# Magnetic behavior of the spin-chain compounds, $Ca_3CuIrO_6$ and $Ca_3CuRhO_6$

S. Rayaprol and E.V. Sampathkumaran[*]

*Tata Institute of Fundamental Research, Homi Bhabha Road, Colaba, Mumbai – 400005, India.*

**Abstract**

The spin-chain compounds, $Ca_3CuIrO_6$ and $Ca_3CuRhO_6$, crystallizing in a $K_4CdCl_6$-derived monoclinic structure, are investigated by ac and dc magnetization, isothermal remnant magnetization as well as heat capacity measurements. The results reveal the existence of a magnetic ordering in the vicinity of 15 K for both the compounds, but the transition appears to be of a complex nature. The existence of a spin-glass component is strongly indicated by the results. We propose that topological effects play a role on magnetism of these compounds. The magnetic properties for these two compounds are interestingly similar as though isoelectronic chemical substitution at the octahedral coordination site does not significantly interfere in the magnetic exchange process.

**I. Introduction**

The spin-chain compounds belonging to a family of the type, $(Ca, Sr)_3ABO_6$ (A,B= metallic, magnetic ions) [1-3], crystallizing in a $K_4CdCl_6$-derived rhombohedral structure, with a triangular arrangement of spin-chains in the basal plane, have started attracting a lot of attention in recent years (see Refs. 4-25 and articles cited therein). In particular, the spin-glass features observed for some of the compounds with Co at the A/B site [11-14, 17-19] (e.g., an unusual frequency ($\nu$) and the magnetic-field (H) dependence of the peaks in ac susceptibility ($\chi$)) are uncharacteristic of canonical spin-glasses and therefore the magnetism of these compounds is apparently quite exotic. The Cu containing compounds present additional complications due to a lattice distortion (mostly monoclinic) [Refs. 4-8] arising out of the Jahn-Teller effect associated with the $3d^9$-configuration of $Cu^{2+}$ ion (A-site) and the ac $\chi$ behavior due to spin-glass freezing is found to be somewhat different [15,16] from those of Co-based systems. It should be noted that, for a Cu compound [25], undergoing triclinic distortion due to a partial site-exchange, long-range antiferromagnetic order (instead of spin-glass freezing) has been reported at low temperatures, despite various unfavorable factors like disorder and multiplicity of bond-distances. It is therefore clear that the spin-glass behavior due to the topological frustration is not a general rule [also see, Refs. 2, 16] within this family, thereby emphasizing the role of other factors, e.g., second and higher nearest neighbor interactions. In view of such interesting situations these compounds present, as a part of our efforts on Cu-containing compounds, we have investigated the compounds, $Ca_3CuIrO_6$ and $Ca_3CuRhO_6$, for their magnetic behavior. To our knowledge, there is no further article on these compounds following the report on the synthesis of the former material [6] and on the dc magnetization (M) behavior of the latter [7,8].

**II. Experimental details**

The compounds in the polycrystalline form were prepared by solid state reaction method. Requisite amounts of high purity (>99.9%) starting components ($CaCO_3$, Ir, Rh, powder and CuO) were thoroughly ground in acetone and heated in Pt crucibles between 800 $^0$C to 1100 $^0$C for 1 week with intermediate grindings. The samples thus formed were found to be single phase. All the lines in the x-ray diffraction patterns (Fig. 1) could be indexed on the basis of monoclinically distorted $K_4CdCl_6$-structure (space group C2/c). While the isothermal M measurements at 1.8 K were performed employing a commercial (Oxford Instruments) vibrating sample magnetometer, the temperature (T) dependent ac and dc magnetic χ behavior were tracked by a superconducting quantum interference device (Quantum Design).

---

[*]Email: sampath@tifr.res.in

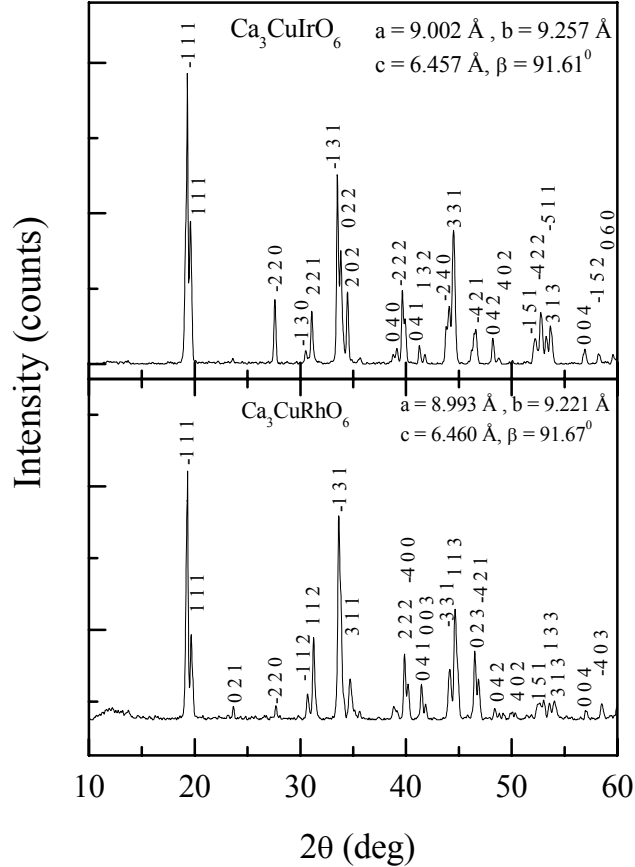

**Fig. 1**     X-ray diffraction patterns (Cu $K_α$) of $Ca_3CuIrO_6$ and $Ca_3CuRhO_6$.

The heat-capacity (C) data in the absence and in the presence of magnetic fields (H) were obtained below 35 K by the relaxation method employing a Physical Property Measurements System (Quantum Design).

**III. Results and discussion**

The results of dc χ measurements measured in the presence of an H of 5 kOe after zero-field-cooling (ZFC) to 1.8 K are shown in figure 2 for both the samples. χ obeys Curie-Weiss law above 150 K. The values of the effective moment ($μ_{eff}$) and the paramagnetic Curie temperature ($θ_p$) obtained from the linear region are: 2.1 $μ_B$ and –12 K for $Ca_3CuIrO_6$ and 2.44 $μ_B$ and –140 K for $Ca_3CuRhO_6$ respectively. As the T is lowered further, there is a gradual deviation from the high-T Curie-Weiss behavior with a pronounced anomaly due to the onset of magnetic ordering well below 20 K.

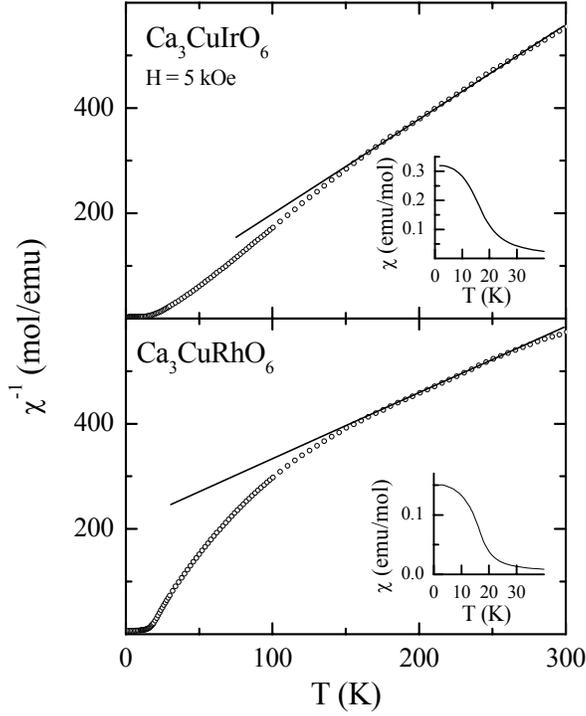

**Fig. 2**  Inverse susceptibility of $Ca_3CuIrO_6$ and $Ca_3CuRhO_6$ as a function of temperature obtained in a magnetic field of 5 kOe (for the zero-field-cooled state of the specimens). A line is drawn through the high temperature linear region. Inset shows the susceptibility data below 30 K in an expanded form.

This finding is in qualitative agreement with the reports of Stitzer et al [7] and Davis et al [8]; however, these authors did not find the theoretically predicted value of 2.45 $\mu_B$ for $\mu_{eff}$ for the spin-only moment of $Rh^{4+}$ (low-spin for the octahedral coordination) and $Cu^{2+}$, whereas our value is in perfect agreement with this combination of electronic structure. However, the observed value for $Ca_3CuIrO_6$ is marginally lower (for $Cu^{2+}$ and low-spin-$Ir^{4+}$); such discrepancies sometimes observed [26] for this family of compounds in the literature have been attributed to anisotropy introduced in some specimens due to quasi-one-dimensional nature of these materials.

We have also taken the dc $\chi(T)$ data below 30 K for H= 100 Oe for the ZFC and FC conditions of the specimen. The plots shown in figure 3 clearly reveal that there is a sudden upturn below about 16 K with a significant bifurcation of ZFC-FC curves in the range 10-15 K. The ZFC-curves [27] exhibit a peak at about 7 and 12 K for the two compounds respectively, while FC-curves tend towards saturation at lower temperatures. The isothermal M curves at 1.8 K shown in figure 4 exhibit an abrupt increase around 300 Oe; the sudden increase of M setting in only after a certain value of H typical of a meta-magnetic transition implies the existence of an antiferromagnetic coupling. M beyond 10 kOe varies sluggishly (that is, without exhibiting saturation even at fields as high as 120 kOe), with a much smaller value of the magnetic moment even at 120 kOe compared to the ferromagnetic spin-only value of 2 $\mu_B$.

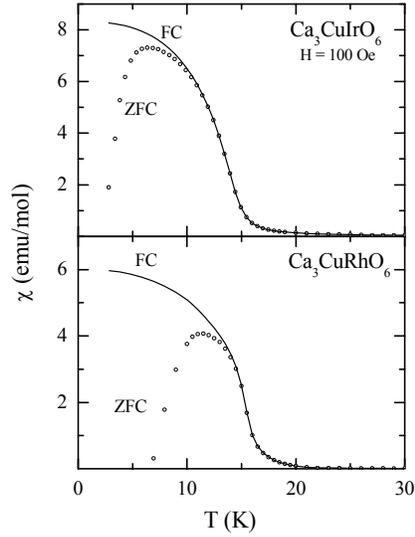

**Fig. 3**     Magnetic susceptibility as a function of temperature for $Ca_3CuIrO_6$ and $Ca_3CuRhO_6$ for the zero-field-cooled (ZFC) and field-cooled (FC) conditions of the specimens for H= 100 Oe.

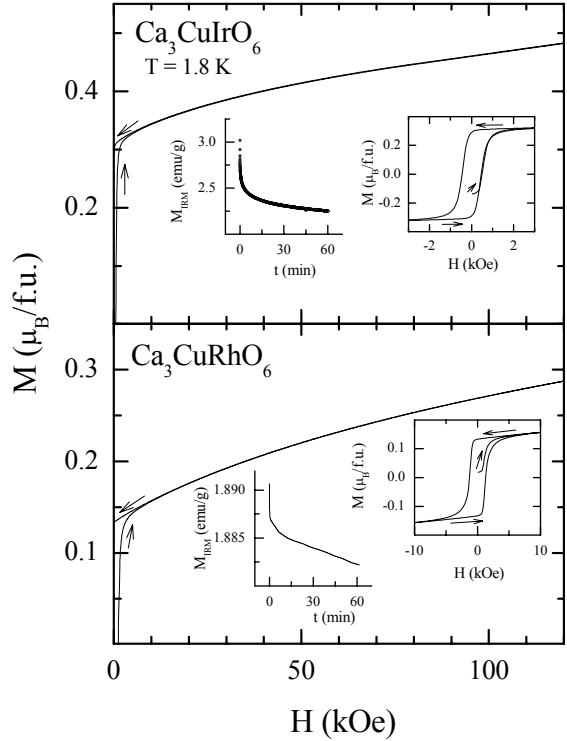

**Fig. 4**     Isothermal magnetization (in the units of magnetic moment per formula unit) as a function of magnetic field at 1.8 K for $Ca_3CuIrO_6$ and $Ca_3CuRhO_6$. Low-field hysteretic behavior is shown in the insets. The arrows mark the way the magnetic field is varied. The time-dependence of isothermal remanent magnetization at 1.8 K is also plotted in the insets.

While these findings support antiferromagnetic ordering (in zero magnetic field), the hysteretic nature of the M-H curves at low fields at 1.8 K (see insets of figure 4) appears to favor ferromagnetic interaction as well. Thus, these M data reveal the existence of both ferromagnetic as well as antiferromagnetic couplings. The question therefore arises whether this inference viewed together with the bifurcation of ZFC-FC dc $\chi(T)$ curves implies spin-glass freezing, which will be addressed below further.

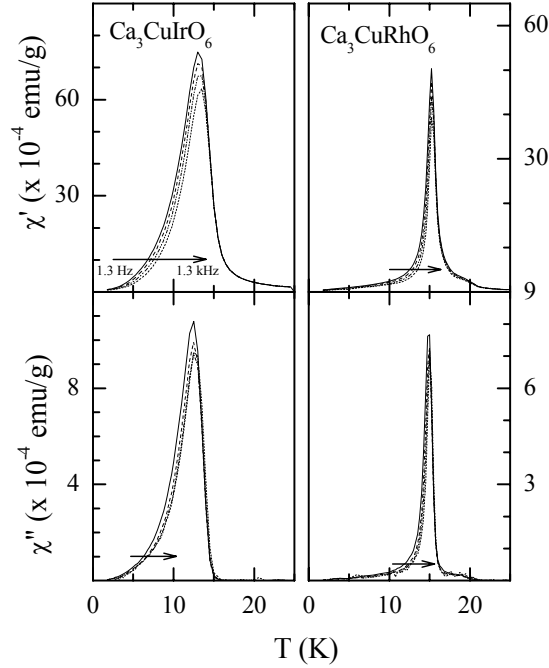

**Fig. 5**  Ac susceptibility (amplitude = 1 Oe) as a function of temperature for $Ca_3CuIrO_6$ and $Ca_3CuRhO_6$ at various frequencies. The curves move with increasing frequency (1.3, 13, 130, 1300 Hz) in the direction of arrows.

We now present the $\nu$-dependent ac $\chi(T)$ behavior (figure 5). It is clear that, in the absence of an external dc H, the real ($\chi'$) part for $\nu$= 1.3 Hz exhibits a well-defined peak near the transition, say at 13 K and 15.3 K, for $Ca_3CuIrO_6$ and $Ca_3CuRhO_6$ respectively. This peak apparently shifts to a marginally higher temperature, say to 13.5 K for $\nu$= 1.3 kHz for the former; though this shift in the peak position is not transparent for the latter, the fact that the left-side of the peak moves to a higher temperature with increasing $\nu$ implies a $\nu$-dependence of ac $\chi(T)$ curves. The shift in the peak temperature (if called as "spin-freezing temperature, $T_f$"), for instance, for the former, corresponds to a value of about 0.01 for the factor [28] $\Delta T_f/T_f\Delta(\ln\nu)$; the features in the imaginary part ($\chi''$) of $\chi$ are quite prominent in the vicinity of $T_f$; the features in ac $\chi(T)$ get wiped out in the presence of an external dc H, say 10 kOe; these properties are typical of canonical spin-glasses.

We have also performed time-dependent isothermal remanent magnetization ($M_{IRM}$) measurements at 1.8 K. For this purpose, after attaining 1.8 K by zero-field-cooling, a magnetic field of 5 kOe was switched on for a time (*t*) of 5 mins, followed by measuring *t*-dependence of M after H was reduced to zero. It is found that, for the same measurements above 15 K, M falls to negligibly small values within few seconds after H is switched off (thereafter remaining constant), while at 1.8 K, the values are found to be large. $M_{IRM}$ is found to undergo a slow decay with *t* (see the insets in Fig. 4), which indicates the existence of a spin-glass-like phase in the sample. While the functional form is logarithmic ($M_{IRM}$= 2.58−0.08$ln$t) for the Ir sample, the one for the Rh sample appears to be more complex. It is obvious from the insets that the net decay over a period of 1 hour is in fact much smaller for the latter compared to the former, and it is not clear whether it implies different degree of complexity of the magnetic phase, e.g., with different fractions of the magnetic ions undergoing spin-glass freezing.

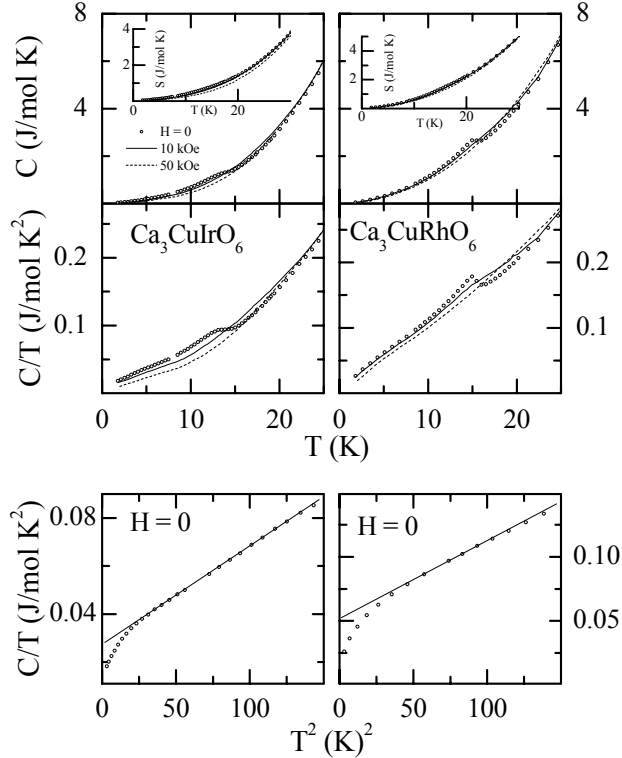

**Fig. 6** Heat-capacity (C) and C/T as a function of temperature for $Ca_3CuIrO_6$ (left columns) and $Ca_3CuRhO_6$ (right columns) in the absence and in the presence of magnetic fields (10 and 50 kOe). The temperature dependence of total entropy is also shown in the insets. The plots of C/T vs $T^2$ are also shown and the continuous lines are drawn to highlight the linear region below 12 K.

The C(T) behavior is shown in figure 6 in various ways. It is apparent that there is a shoulder (very weak peak) in the plots around 13 K (15 K) for $Ca_3CuIrO_6$ ($Ca_3CuRhO_6$), not only in C(T) but also in C/T(T). Thus, the entropy change in the range 13-15 K at the transition (Fig. 6, inset) is negligible. In other words, the absence of a *prominent* λ-peak in C(T) raises a question whether a well-defined magnetic structure, if present, proceeds through entire crystallite, in other words, whether the magnetic transition is of an inhomogeneous (or broadened) type. Spin-glass-like freezing could be one source of this type of transition. In support of this, the plot of C/T versus $T^2$ (see Fig. 6) is linear below about 12 K (down to 4.5 K) with a large linear term, typical of spin-glasses. It is however to be noted that, though the anomaly in C at the respective transition temperatures is weak, there is a sharp peak (see Fig. 6), visible at least for $Ca_3CuRhO_6$. This could imply that there is a finite fraction of magnetic ions undergoing long-range magnetic ordering with a well-defined magnetic structure centering around 15 K. It is therefore not clear at present whether one has to invoke the idea of 'partially-disordered antiferromagnetic structure (PDA)' advanced for $Ca_3Co_2O_6$ and $Ca_3CoRhO_6$ [Refs. 9,11,13], though we can not rule out possible formation of clusters of antiferromagnetic segments created by the defects along the

66

chains [15], coupling among themselves like a cluster spin-glass. We have also measured C in the presence of 10 and 50 kOe and we find that the kink around the transition is smeared out. The insets in figure 6 reveal how the applied field modifies the way the total entropy evolves. The S(T) curves, say for H= 0 and 50 kOe, tend to overlap at a temperature far above $T_f$, which appear to favor the proposal of magnetic inhomogeneity persisting over a wide temperature range above $T_f$. Incidentally, these correlations could possibly extend to temperatures as high as 150 K, as inferred from the deviation of $\chi(T)$ from the high-T Curie-Weiss behavior (see Fig. 1).

Finally, we would like to add that there is a weak change in the slope of C/T vs. $T^2$ plots around 4.5 K in both the samples, which indicates further complexities of the magnetic phase of these compounds at such temperatures. The linear region below 4.5 K yields the values of linear coefficient and Debye temperature as: 15 mJ/mol $K^2$ (21 mJ/mol $K^2$) and 275 K (225 K) for $Ca_3CuIrO_6$ ($Ca_3CuRhO_6$) respectively.

**IV. Summary**

The results presented in this article suggest that the two compounds, $Ca_3CuIrO_6$ and $Ca_3CuRhO_6$, exhibit complex magnetic behavior with the onset of magnetic ordering around 15 K. Several experimental results viewed together (the bifurcation of ZFC-FC $\chi(T)$ curves, ν-dependence of ac χ and its response to the application of H, the slow decay of $M_{IRM}$, low-field hysteretic M-H behavior without saturation at high fields, and a broad C(T) anomaly with a large linear term) indicate the existence of a spin-glass-like component. However, a well-defined kink/shoulder, though very weak, appearing in C(T) around 15 K points towards the existence of a well-defined magnetic structure for some fraction of magnetic ions. Thus, the results offer evidence for some kind of random magnetism in the bulk of these samples and this type of magnetism in such stoichiometric materials may have its origin in the topological frustration of coupling among magnetic ions. It is worthwhile to perform careful neutron diffraction studies to understand possible complex magnetic structure including PDA structure.

Finally, a careful look at all the data presented in this article indicates that the properties of both the compounds are apparently similar in many respects. Thus, the chemical substitution at the octahedral site by an isoelectronic ion does not bring about any major qualitative change in the measured physical properties, except perhaps a marginal depression of magnetic ordering temperature by negative chemical pressure. However, a substitution by Ru for this site results in a long range ferri-magnetic ordering [5], *without spin-glass features* [29] in ac χ(T). The fact that the ν-dependence of the peak temperature in χ'(T) of the title compounds falls in line with those of other Cu-based spin-glasses of this family, raises a question whether the unusual spin-glass anomalies in the Co analogues, $Ca_3CoIrO_6$ and $Ca_3CoRhO_6$ [10,11,14,15] are in some fashion decided by Co. Thus, this work enables us to bring out the role of electronic structure of A and B ions on magnetic behavior. We hope these results will be helpful for the theoretical advancement of the field of geometrical frustration.

**Acknowledgements:**

We thank K. K. Iyer and Kausik Sengupta for their valuable help during experiments.